\documentclass[twocolumn,showpacs,preprintnumbers,amsmath,amssymb]{revtex4}
\usepackage{graphicx}
\usepackage{dcolumn}
\usepackage{bm}

\begin{document}

\title{Electronic structure and properties of superconducting materials
with simple Fermi surfaces.}

\author{T. Jarlborg}

\affiliation{
DPMC, University of Geneva, 24 Quai Ernest-Ansermet, CH-1211 Geneva 4,
Switzerland
\\}


\begin{abstract}
The electronic structures of the ground state for several different superconducting materials,
such as cuprates, conventional 3-dimensional superconductors, doped semiconductors
and low-dimensional systems, are quite different and sometimes in contrast to what is
supposed to make a superconductor.  Properties like the Fermi-surface (FS) topology, density-of-states (DOS), stripes,
electron-phonon coupling ($\lambda_{ep}$) and spin fluctuations ($\lambda_{sf}$) are analyzed in order to find clues to
what might be important for the mechanism of superconductivity. A high DOS at $E_F$
is important for standard estimates of $\lambda's$, but it is suggested that superconductivity
can survive a low DOS if the FS is simple enough. Superconducting fluctuations are plausible from coupling 
to long wave length modes in underdoped cuprates, where short coherence length 
is a probable obstacle for long-range superconductivity.
Thermal disorder is recognized as a limiting factor for large $T_C$ independently of doping.

\end{abstract}


\maketitle

\section{Introduction.}

The electronic structures of transition metals (TM) and their alloys and compounds show 
that several d-bands cross the Fermi energy $E_F$. Their Fermi surfaces are complex 
and occupy about all parts of the Brillouin zone (BZ). Calculations of electron-phonon coupling (EPC)
$\lambda_{ep}=NI^2/M\omega^2$, where $N$ is the density-of-states (DOS) at $E_F$,
$M$ an atomic mass and $\omega$ a weighted phonon frequency, became quite popular in the
70-ties when the matrix element $I$ could be determined quite easily from the band structure
by the Rigid Muffin-Tin Approximation (RMTA) \cite{gasp}. 
Then, from BCS \cite{bcs} or the McMillan
formula \cite{mcmill} the superconducting $T_c$ are estimated, and the results for pure elements, TM alloys,
TM nitrides and carbides, C15 and A15 compounds are quite reasonable \cite{papa,pic,daco,c15,arb,jmp,papa2}.
The calculated $T_C$ are not precise, but good superconductors are clearly separated from less good
or bad ones, with the observation that a high $T_C$ needs a high $N$. 
The quantitative agreement with observed $T_C$ is improved when the pair-breaking effect
from spin-fluctations are taken into account through the coupling constant $\lambda_{sf}$ \cite{berk,fay,vanad,daam}.
But, on the other hand,  superconductivity in the very lightly doped semiconductors 
SrTiO$_3$ (STO), WO$_3$, diamond (C) and Si \cite{raub,shoo,suz,lin,diam,bust}
is not easy to understand.
Their DOS are very small and $T_C$ goes to zero already at modest dopings \cite{matt,tjsto,boe,lee,bla,xia},
despite the fact that the DOS normally should go up with doping. 
The problems with
the high-$T_C$ cuprates are well-known \cite{pick}. The band gap in undoped cuprates
is absent in density functional theory (DFT) band calculations. Moreover,
EPC alone is probably too weak to explain $T_C$, since $N$'s are small with few bands at $E_F$ and
simple FS's. Spin-phonon coupling (SPC) enforces $\lambda$'s \cite{tj1,tj7}, 
and spin-fluctuations could be determining for $T_C$  \cite{pines},
as also could be the case for low $T_C$'s in ZrZn$_2$ \cite{zrzn2} and Fe under high pressure \cite{shim,jac,tjfe}.

Here we make an effort to understand the difficult cases by doing some corrections of how EPC is to be evaluated
for materials with simple FS.


\section{Results and discussion.}

\subsection{Bands and coupling constants.}

The band structure results presented here are made using the linear muffin-tin orbital (LMTO) method \cite{lmto,bdj} and the
local spin-density approximation (LSDA) \cite{lsda}.  Electronic correlation beyond that contained within
 Density-Functional Theory (DFT) is disregarded.
It is sometimes argued that correlation is too strong for having traditional bands in undoped cuprates
when the d-band is half filled. But DFT bands and FS's for doped cuprates agree well with ARPES 
(angular-resolved photoemission
spectroscopy) and ACAR (angular correlation of positron annihilation radiation)  \cite{pick,dama,posi}. 
The calculation of $I$ in the RMTA leads to dipole transitions that
couple states $\ell$ to $\ell \pm 1$. The d-bands in TM compounds are hybridized with p and f, and this fact
makes $I_{RMTA}$ large.  From its name it is understood that
the potential is displaced "rigidly" to get the dipolar matrix element as a first order
change of the band energy as function of the displacement. However, in ionic systems there are also changes
in the local Madelung potential when an atom is displaced, which leads to a monopolar matrix element \cite{ssc88}.
This makes the coupling largest for unhybridized bands, as for the Cu-d band in cuprates, the d-band
in doped SrTiO$_3$ (STO) and WO$_3$, and the p-band in hole doped C and Si. Since $I_q$ (at 
k-points on the FS) is the first order
change of the band energy $\epsilon_k$, it can also be obtained from the change in band energy when a phonon
modifies the potential. The advantage with this method is that both the monopole and dipole contribution are 
included.  The result also takes screening into account when the calculations are selfconsistent. 
This method has been used in
the nearly-free electron (NFE) model \cite{tj6,tj9}, as well as in
fully self-consistent LMTO calculations \cite{tj7},
to show that the largest $I_q$ appears for phonon q-vectors that span the FS. Only few q-vectors generate
gaps at $E_F$, and together with the dominant contribution from nested FS it leads to involvement of
rather few q-modes for simple FS \cite{ssc11,ssc14}. 

\subsection{Doped semiconductors.}

\begin{figure}
\includegraphics[height=7.0cm,width=8.0cm]{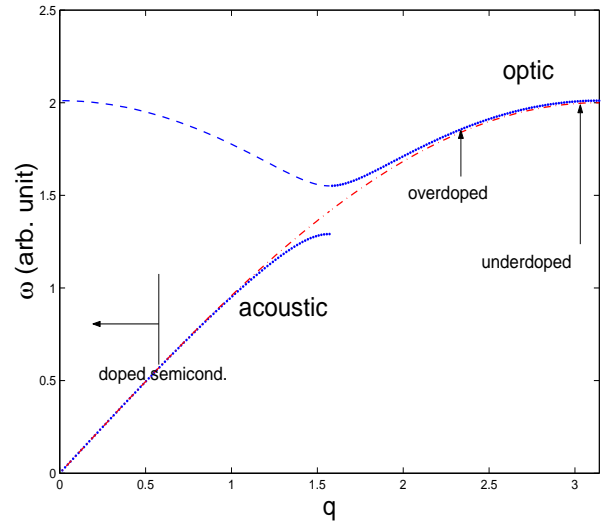}
\caption{(Color online) Schematic phonon dispersion along $k_x$.
It is argued that the largest EPC in doped semiconductors is found for 
the acoustic low energy mode.  Optic modes in underdoped cuprates
are not efficient for long-range superconductivity, but acoustic modes grow
in importance at higher doping.
}
\label{figphon}
\end{figure}

The FS's in electron doped STO and hole doped
C and Si, consist of simple small pockets at the $\Gamma$-points, so that all of the electron-lattice
interactions are concentrated to a small fraction of the BZ near the zone center. 
The logarithm of the BCS equation for $T_C$; 
$M\omega_q^2 = N_qI_q^2 log(1.13\hbar\omega/k_BT_C)$, permits a separation of 
the dominant energy cost from excitations 
of phonons with vector $q \sim k_F$ from the electronic energy gain from different pieces of the FS at $N_q$. 
An average of $\omega^2$ over the entire phonon spectrum
includes frequencies up to the Debye frequency for q-vectors at the limit of the BZ, but only
up to a much smaller cutoff for weakly doped semiconductors, since no gain
in total energy comes from larger $q$, see Fig. \ref{figphon}. 
The cost of phonon energy is therefore
much reduced, and it compensates for the low $N$. A simple estimation of these energies for a
Debye phonon DOS, $F(\omega)$, and free electron DOS (both in 3-dimensions) 
at low doping show also that this energy balance
becomes less favorable at higher doping, so that $T_C$ goes down \cite{ssc14}. 
Thus, the absence of large-q phonon excitations 
can explain superconductivity from acoustic small-q, low-energy phonons in weakly
doped semiconductors despite their low DOS, and the drop in $T_C$ for increased doping \cite{ssc14}.
According to this, only long wavelength phonons appear at $T=0$ together with
lattice disorder from zero-point motion (ZPM).

\subsection{Cuprates.}

The FS in cuprates involves large-q vectors between nested sections of the 2-dimensional circular FS. The FS becomes
almost diamond shaped at optimal doping, when the FS reaches the X-point of the BZ and the DOS
has a van-Hove singularity peak. This permits
reduced number (ideally only two) of phonon/spin mode excitations between electron states on two
parallel FS sheets in order to create a gap over the whole FS.
This not exactly as the small-q situation in doped semiconductors, but the 2-dimensional
structure with few flat sections of the FS is likewise favorable to superconductivity
despite a low $N$ \cite{ssc14}. 
In this context it can be noted that
$T_C$ in Ba$_2$CuO$_3$ is found to be much larger than in La$_2$CuO$_4$ \cite{liu,jin,geba}.
The electronic interplane interaction is reduced in the former material because of less
apical oxygens, and it makes the FS even more flat at optimal doping than in  La$_2$CuO$_4$
\cite{jbmb,jbb,ssc14}.
Additional ordering of dopants into static stripes has been suggested to lead to a segmentation of the FS
and enhanced $T_C$ \cite{jb,chm,poc}. It is not clear if such enhancement can come from fewer phonon/spin
excitations or from increased DOS because of stripe order \cite{apl}.
It can also be noted that weak signs of ferromagnetism are present at high doping \cite{bj,son}.
However, the doping is very high and beyond the concentration where superconductivity can be found. 
Anti-ferromagnetic (AFM) order is installed on Cu at zero doping. The amplitude of AFM moments and
local exchange splittings decreases for increasing doping, which introduces a doping dependence
of the coupling constant for spin fluctuations ($\lambda_{sf}$). Calculations of the strength of $\lambda_{sf}$
are uncertain because of the problem with LDA to describe the static AFM order and the band gap for undoped
cuprates \cite{tj4}, but as expected, there is a clear decrease of $\lambda_{sf}$ as function of doping $h$ 
(holes/Cu) \cite{tj6}.  In addition, phonons and spin waves are mutually enforcing each other
through SPC \cite{tj1,tj7}.

\subsection{Disorder and fluctuations in cuprates.}

\begin{figure}
\includegraphics[height=6.0cm,width=8cm]{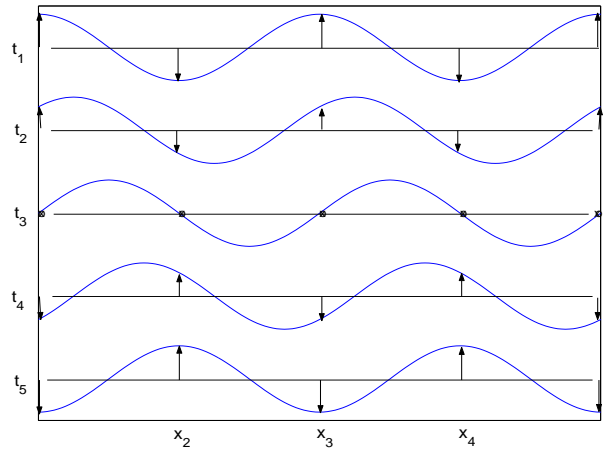}
\caption{(Color online)  Time evolution of a pure optic phonon/spin wave mode
for 5 atoms along $x$. The arrows indicate atomic displacements and/or magnetic moments
for each atomic site.
Note that all distortions/moments are zero at time $t_3$.
}
\label{figop}
\end{figure}

\begin{figure}
\includegraphics[height=6.0cm,width=8cm]{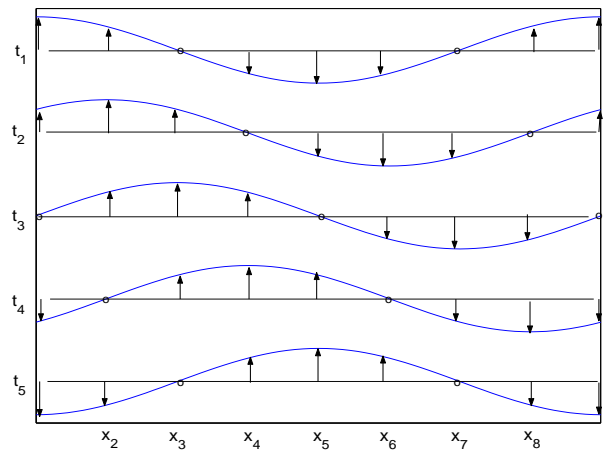}
\caption{(Color online) Time evolution of an acoustic phonon for 
9 atoms along $x$, the arrows indicate atomic displacements.
Note that the wave is identical at all times except for the phase. 
A cuprate spin wave would have AFM spin orientation on near neighbors, as in Fig. \ref{figop}, but
all modulated by the long-range envelope function.
}
\label{figac}
\end{figure}

The anti-ferromagnetic (AFM) spin arrangement on Cu sites along $x$ in undoped cuprates can be described by 
$V_Qexp(-iQx)$, where $V_Q$ is the exchange splitting and $Q=\pi/a_0$. Doping makes the gap to appear at
lower energy and the band ($\epsilon_k$) is crossing $E_F$ for $k=Q-q$. The AFM order is modulated by
$exp(iqx)$ and the spin potential becomes 
\begin{equation}
V(x) = V_{Q-q}exp(-i(Q-q)x))
\label{eqV}
\end{equation}
 Optimal doping corresponds to $q \approx Q/4$,
with "stripes" covering 4 sites, and the exchange splitting decreases for larger $q$ \cite{tj7,tj6}.
The vector $q$ is small at very low doping, and the spin arrangement is almost like an optic wave, see
Fig \ref{figop}. The spin
distribution of an optic wave (or atomic displacements for phonons) is regularly zero over a large region
(at times $t_3$ in Fig. \ref{figop}), which in principle makes $\lambda$ time dependent. 
An acoustic wave, however, has always the same shape (Fig. \ref{figac}).
 If $q$ is very small (almost optic), with very wide 
stripes covering several tens of atoms, then the wave is
intact as for a short acoustic wave. But rather wide regions between the stripes have very small 
spin/distortion amplitudes (for the short
wave in Fig. \ref{figac} one of 4 sites has zero spin/distortion). The suggestion is that
long-range superconductivity is not possible if the regions with small spin/distortion
amplitudes are wider than the coherence length. Alternatively, superconductivity could still
be possible, but at a lower $T_C$ than what is expected from the large amplitudes within the stripes.
Possible short-range superconducting fluctuations within
the stripes are sufficient for a high $T_C$, and can be the cause of the 
pseudogap at $T^*$. 

Some numbers to give substance to this idea: The BCS coherence length, $\xi = 2 \hbar v_F / \pi \Delta$,
where $v_F$ is the Fermi velocity (taken from the calculated bands in La$_2$CuO$_4$ \cite{tj11} to be
between 10 and 35 $\cdot 10^4 m/s$ 
from underdoping, UD, near $h$=0.05 to optimal doping, OD, near h=0.15) 
and $\Delta$ the superconducting gap (here $2\Delta = 3.5 T_C$, 
and $T_C = T^*$ is supposed to be due to superconducting fluctuations. Typical $T^*$ are
in the range 60-30 $meV$ from UD to OD \cite{dama}) from which $\xi$ can be estimated to be 
between 1$a_0$ and 7$a_0$ at low T, respectively. 
This gives very approximately; $\xi \sim 30h$ (units of $a_0$).
The wave lengths, $\Lambda$, of spin waves and phonons are related to doping, $h$ holes/Cu. From
LMTO supercell calculations $\Lambda = 2/h$ in units of $a_0$ ($\Lambda$ is the length of a unit cell
along [100] in which one spin wave or two phonon waves can fit), and the amplitudes $V_{(Q-q)}$
are calculated to be of the order 25-17 mRy between UD and OD \cite{tj7}. In the wave drawn in Fig \ref{figac}
the zero spin/displacement region covers 1/8th of the cell. By assuming that $\xi$ must be
larger than the width $x_c$ in which $V(x)$ is smaller than $\sim 0.25$ of the maximal $V_{(Q-q)}$ we
get the condition $x_c < arcsin(0.25)/\pi/h$, which gives $x_c$ smaller than 1.6 to 0.5 $a_0$ from
UD to OD.
 Thus, $\xi > x_c$ for $h$ smaller than about 0.07, for which there should be no long-range superconductivity according to this
reasoning.

\begin{figure}
\includegraphics[height=6.0cm,width=8cm]{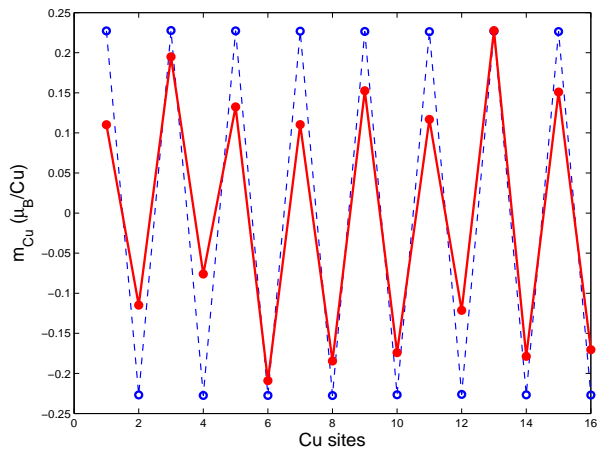}
\caption{(Color online)  Magnetic moments per atom on Cu in ordered (blue open circles connected by
broken lines) and disordered (red circles connected by lines) La$_{32}$Cu$_{16}$O$_{64}$,
 where an applied field of $\pm 8$ mRy sets up
an AFM spin alignment along the {100} direction in the supercell. The disorder corresponds
to that of room temperature.
}
\label{figafm}
\end{figure}

The value of $\xi$ increases when $T$ gets closer to $T_C$, but 
at higher temperature there are thermal disorder of the lattice \cite{san}. The latter is known to induce 
band broadening ($\Delta\epsilon_k$) and temperature
dependent variations of non-superconducting properties in many materials \cite{fesi,fege,dela,bron,cevib}.
There will be irregular variations of the local potential as a function of time,
more so at high temperature and for a soft lattice. 
Well-defined waves along the lattice are required in order to generate a neat gap at $E_F$, when
the Coulomb-part and/or the spin-part of the potential is close to cosine-like shape, as in eq. \ref{eqV}.
Thermal disorder generates random potential fluctuations on the atomic sites, and if the fluctuation amplitudes,
$V_T$, are comparable to $V_{(Q-q)}$ then they will interfere with the cosine wave and finally make it unrecognizable.
Cuprates are unique because of the AFM order on Cu, and  
the spin-polarized part of $V_T$ turns out to be more sensitive to disorder than the Coulomb part.
Disorder makes large shifts of the local Madelung potential, but the exchange splitting ($\chi_{Cu}$)
on different Cu sites 
develops quite differently, and the calculation show that $V_T$ at RT for a disordred supercell of
La$_{32}$Cu$_{16}$O$_{64}$ is larger than $V_{(Q-q)}$ calculated for 
a cosine SPC wave in a Hg based cuprate \cite{tj4}.
Calculations of $V_{(Q-q)}$ for simple AFM in undoped La$_{32}$Cu$_{16}$O$_{64}$ show that the average of moments
and  $V_{(Q-q)}$ goes down with disorder. At RT the averages are approximately 2/3 of those for the ordered structure,
and the amplitude of individual moments varies by a factor of three (or more for weak spin waves), 
which shows that AFM waves are much perturbed
by disorder,
see Fig \ref{figafm}. The average valence electron charge on Cu increases from 10.39 to 10.47 at this level of disorder,
as if disorder reduced the level of hole doping.
Band broadening is proportional to the rms-average of $V_T$, and long-range
superconductivity could be suppressed by thermal disorder if $V_T$
exceeds the superconducting energy gap. The band broadening from thermal
disorder is difficult to calculate, since it needs large supercells. In other materials the averaged broadening
$\Delta \epsilon$ is calculated to be
of the order 50 meV at room temperature (RT) when the average of the atomic displacements is about  
0.1~\AA \cite{fesi,bron}. Here, for undoped LCO $\Delta\epsilon_k$ is about 28 meV in the
Coulomb part, and about 60 meV when the calculations are made for an AFM configuration.
Potential shifts on Cu caused by disorder are typically larger than $\Delta \epsilon$ on about 1/4 of 
all Cu. 

\begin{figure}
\includegraphics[height=6.0cm,width=8cm]{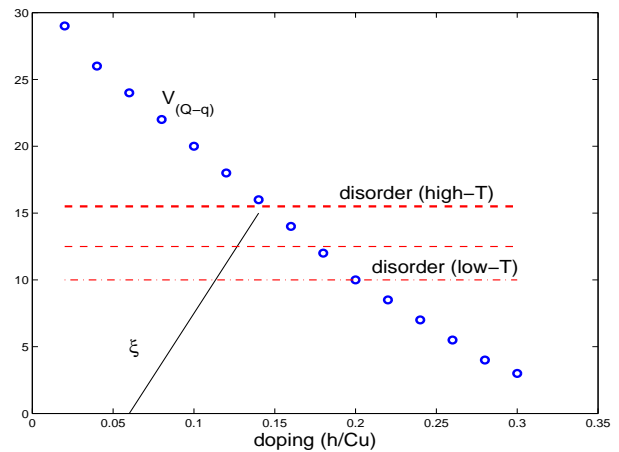}
\caption{(Color online) A sketch of a phase diagram for cuprates. The blue circles 
indicate the coupling strength $V_{(Q-q)}$, but
$T_C$ is zero to the left of the line for $\xi$ due to a short coherence length. Thermal disorder will gradually
destroy well-defined waves at a certain level indicated by the red horizontal lines.
Long-range superconductivity
would be possible below the lines, but only fluctuations on the left side of the $\xi$-line.}
\label{phdiagfig}
\end{figure}

The coupling strength $V_{(Q-q)}$ for spin fluctuations increases when $h \rightarrow 0$, because the AFM state 
is more and more developed at low doping and finally the cuprates become insulators with stable AFM moments
at zero doping. But $T_C$ is known to have a maximum for $h \approx$ 0.15. 
In a very approximate drawing of a phase diagram
in Fig. \ref{phdiagfig}  we suggest that superconducting fluctuations can survive at low doping up to large
temperature proportional to $V_{(Q-q)}$. No fluctuations appear above $T_C$ at high doping, because $V_{(Q-q)}$
is not large. Thermal disorders supress superconductivity at all dopings.

If disorder puts a limit to the long-range superconductivity, one may search for ways to recuperate a higher $T_C$
by some kind of stabilization of superconducting fluctuations. At low $T$ it seems that longer $\xi$ can be obtained
via higher $v_F$, as probably is achieved naturally in many cuprates by a static pseudogap. This would permit
an extended pairing at UD, but superconductivity will be weaker because a higher $v_F$ implies a lower
$N(E_F)$ and weaker coupling.  
Anharmonicity could also be favorable, since the thickness of a zone with low $V_{(Q-q)}$ will
be narrower for a step-like wave than for a cosine wave.
At higher $T$ it is difficult to get rid of thermal disorder. Higher coupling
constants, especially for spin fluctuations, should be efficient to enforce clean spin waves, but the
problem is that also $V_T$ will profit from high exchange enhancement. Therefore, bad influence from
disorder could spoil the effect of enforced spin waves. Structural disorder will be smaller in a stiffer lattice
at a given $T$, and since the spin disorder is largely an effect of the structural disorder, it suggests that
lattice hardening would help. This can be achieved by applying pressure, but the strength of the spin wave
($V_{(Q-q)}$) must not decrease when the volume is reduced.  The idea that a higher $N(E_F)$ provides a 
larger gain in electronic energy from a superconducting gap, and hence stronger coupling constants
and higher $T_C$, is still valid at large disorder. How to modify the cuprate band structures in order to
get higher DOS without destroying other properties might be difficult, perhaps ordering of impurities
into stripes is a promising way \cite{jb}.

\section{Conclusion.}

Electron-phonon interaction and coupling from spin fluctuations are largest when
the q-vector of the lattice/spin wave is equal to the k-vector on the FS, at least
for the cases of simple FS's of cuprates and some doped semiconductors. This is
evident for NFE bands, but it is also seen in the ab-initio calculated band structures of
cuprates \cite{tj7}. For a simple FS it is therefore possible that superconductivity relies only 
on a few phonon/spin excitations, which makes superconductivity
possible even if $N(E_F)$ is small. 
In contrast, for a complicated FS practically
the entire phonon/spin spectrum is needed to open gaps over all parts of the FS.
Further, it can be noted that a pure optic mode should have a frequency dependent coupling.
A long wavelength modulation of an optic mode, as for underdoped cuprates, behaves as a
normal acoustic mode, but wide regions with small coupling could prevent long-range
superconductivity. Shorter wave lengths of acoustic modes at higher doping do not have
this property, but thermal disorder at high temperature will cause potential fluctuations
which compete with and prevent long-range superconductivity. Based on such mechanisms
we speculate that the disorder  prevents long-range superconductivity,
but that superconducting fluctuations are left behind and they will cause a pseudogap
below $T^*$. There are no obvious solutions for an easy transformation of superconducting fluctuation into
long-range superconductivity with higher $T_C$, but the sensitivity of $T_C$ due to varied conditions 
is expected to be different at low and high doping.

\end{document}